\documentclass[pss,fleqn]{w-art}
\usepackage{times}
\usepackage{amssymb}
\usepackage{w-thm}
\usepackage{graphicx}

\newcommand*{\ep}{\epsilon}
\newcommand*{\de}{\delta}

\newcommand*{\si}{\sigma}

\newcommand*{\ua}{\uparrow}
\newcommand*{\da}{\downarrow}
\newcommand*{\eff}{\mathrm{eff}}

\newcommand*{\abs}[1]{\left|#1\right|}
\newcommand*{\aver}[1]{\left<#1\right>}
\newcommand*{\averaa}[1]{\langle #1\rangle}

\begin{document}
\DOIsuffix{theDOIsuffix}
\Volume{XX}
\Issue{1}
\Month{01}
\Year{2003}
\pagespan{1}{}
\Receiveddate{}
\Reviseddate{}
\Accepteddate{}
\Dateposted{}


\title[ELECTRONIC STRUCTURE AND PARITY EFFECTS ...]{%
  ELECTRONIC STRUCTURE AND PARITY EFFECTS IN CORRELATED NANOSYSTEMS
}


\author[A.\ Rycerz]{Adam Rycerz\footnote{Corresponding
    author, e-mail: {\sf adamr@th.if.uj.edu.pl}\\ 
    $\mbox{ }$ \hspace{5mm}
    Present address: Instituut--Lorentz, Universiteit Leiden,
    P.O.\ Box 9506, NL--2300 RA Leiden, The Netherlands}\inst{1}} 
\address[\inst{1}]{Marian Smoluchowski Institute of Physics, 
  Jagiellonian University, Reymonta~4, 30--059 Krak\'{o}w, Poland}
\author[J.\ Spa{\l}ek]{Jozef Spa{\l}ek\inst{1}}

\begin{abstract}
We discuss the spectral, transport and magnetic properties of quantum 
nanowires composed of $N\leqslant 13$ atoms and containing either even or odd 
numbers of valence electrons. 
In our approach we combine
\textbf{E}xact \textbf{D}iagonalization and \textbf{Ab} \textbf{I}nitio 
calculations (EDABI method). 
The analysis is performed as a function of the interatomic distance. 
The momentum distribution differs drastically for those obtained for even 
$N$ with those for odd $N$, whereas the Drude weight evolve smoothly.
A role of boundary conditions is stressed.

\end{abstract}
\maketitle                   

For low--dimensional systems, 
the procedure starting from the single--particle picture (band 
structure) and including subsequently the interaction via a \emph{local}
potential may not be appropriate. 
In this situation, one resorts to parametrized models of correlated electrons,
where the single--particle and the interaction-induced aspects of the 
electronic states are treated on equal footing.
The single--particle wave--functions are contained in the formal expressions
for model parameters and should be calculated separately. 
We have proposed  \cite{spary} to combine the two efforts in an exact manner.

A separate question concerns the role of boundary conditions (BC\textit{s}) 
in atomic 
rings, particularly under the presence of spin frustration for \emph{odd} 
number of atoms. This problem was investigated numerically \cite{genbc} 
and the optimal BC\textit{s} for a correlated system were found to remain 
usually the same as for the ideal Fermi gas on the lattice. 
However, the general proof of this basic fact have been elaborated very 
recently \cite{nakano}.

In our method of approach (EDABI), we determine \emph{first} rigorously the
energy of interacting particles in terms of the microscopic parameters for
a given BC\textit{s} and then allow the single--particle wave functions 
(contained in the parameters) to relax in the correlated state.
The method has been overviewed in \cite{sparev}, so we concentrate here on its 
application to nanochains  of $N\leqslant 13$ atoms, containing either
\emph{even} or \emph{odd} number of electrons.
The discussion of parity effects complements our recent study of correlated 
nanochains \cite{ryspa}, where we consider the properties of even--$N$ systems 
only.

We consider the system of $N$ lattice sites, each containing a single valence 
orbital and (i.e.\ hydrogenic--like atoms). 
Including \emph{all} long--range Coulomb interaction and neglecting other
terms, one can write down the system Hamiltonian in the form
\begin{equation}
\label{hameff}
  H=\ep_a^\eff\sum_j n_j 
  +t\sum_{j\si}\left(e^{-i\phi/N}c_{j\si}^{\dagger}c_{j+1\si}
  +\mbox{h.c.}\right)
  + U\sum_in_{i\ua}n_{i\da} + \sum_{i<j}K_{ij}{\de n_i}{\de n_j},
\end{equation}
where $\delta n_i\equiv n_i-1$, $\epsilon_a^{\rm eff}=
\epsilon_a+N^{-1}\sum_{i<j}(2/R_{ij}+K_{ij})$ (in Ry) is the effective atomic
level, $R_{ij}$ is the distance between the $i$--th and $j$--th atoms,
$t$ is the nearest--neighbor hopping, $\phi$ is the fictitious (dimensionless)
flux through the ring, 
$U$ and $K_{ij}$ are the intra-- and inter--site Coulomb repulsions.
The last term represents the \emph{correlated}
part of the long--range interaction. One can easily show, that the
unitary transformation $c_{j\si}\rightarrow e^{-i\phi j/N}c_{j\si}$ 
allows to accumulate all the complex phase factors in the terminal hopping
term \cite{poilb}, which than takes the form 
$t(e^{-i\phi}c_{1\si}^{\dagger}c_{N\si}+\mbox{h.c.})$ and can be regarded
as \emph{generalized} BC Such form is particularly convenient for numerical
purposes, since majority of the hopping terms are real. Hereinafter, 
we do not distinguish between the system with a fictitious flux and with
generalized BC\textit{s}.

The Hamiltonian (\ref{hameff}) is diagonalized in the Fock space with 
the help of Davidson technique \cite{davidson}. 
As the microscopic parameters 
$\epsilon_a^{\rm eff}$, $t$, $U$, and $K_{ij}$ are calculated
in the Gaussian basis composing the Wannier functions, the orbital size of
the 1$s$--like state is subsequently
adjusted to obtain the minimal ground state energy $E_G$ as a function of the
interatomic distance $R$. 

\begin{figure}[!t]
\setlength{\unitlength}{0.01\textwidth}
\begin{picture}(100,33)
\put(0,0){\includegraphics[width=0.5\textwidth]{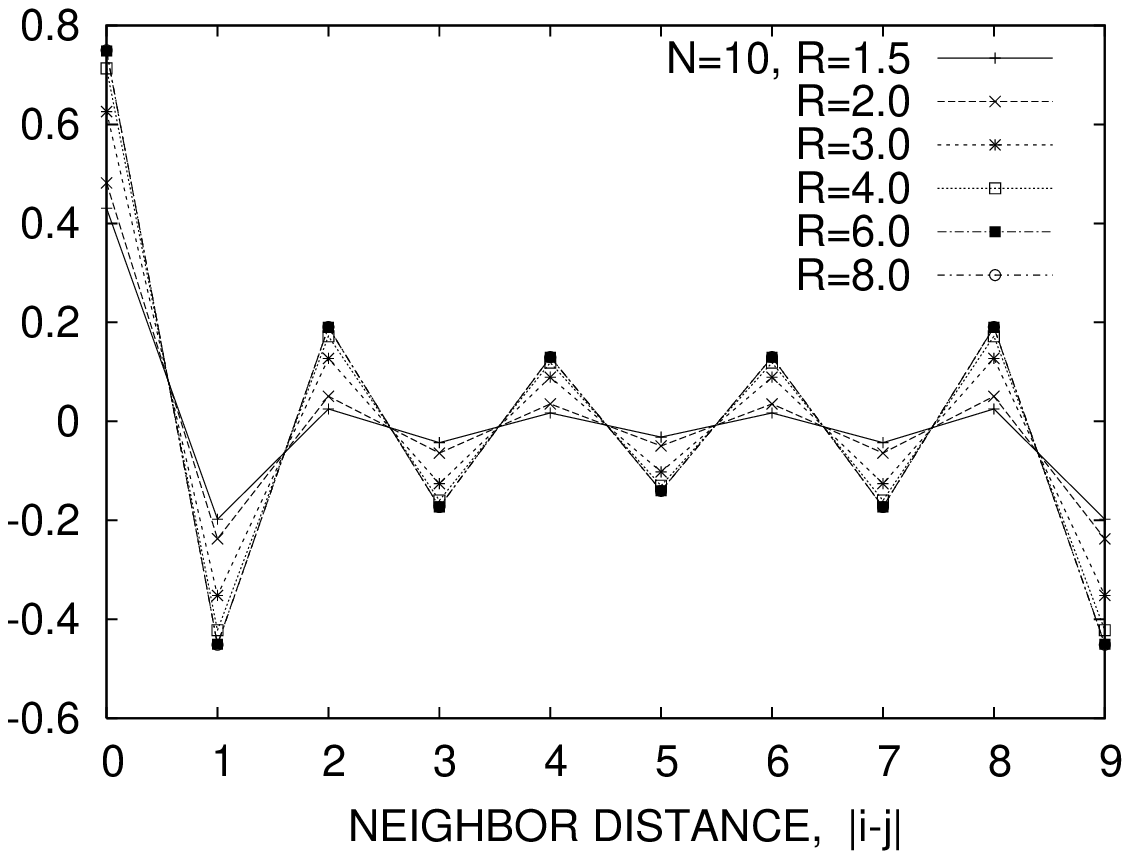}}
\put(50,0){\includegraphics[width=0.5\textwidth]{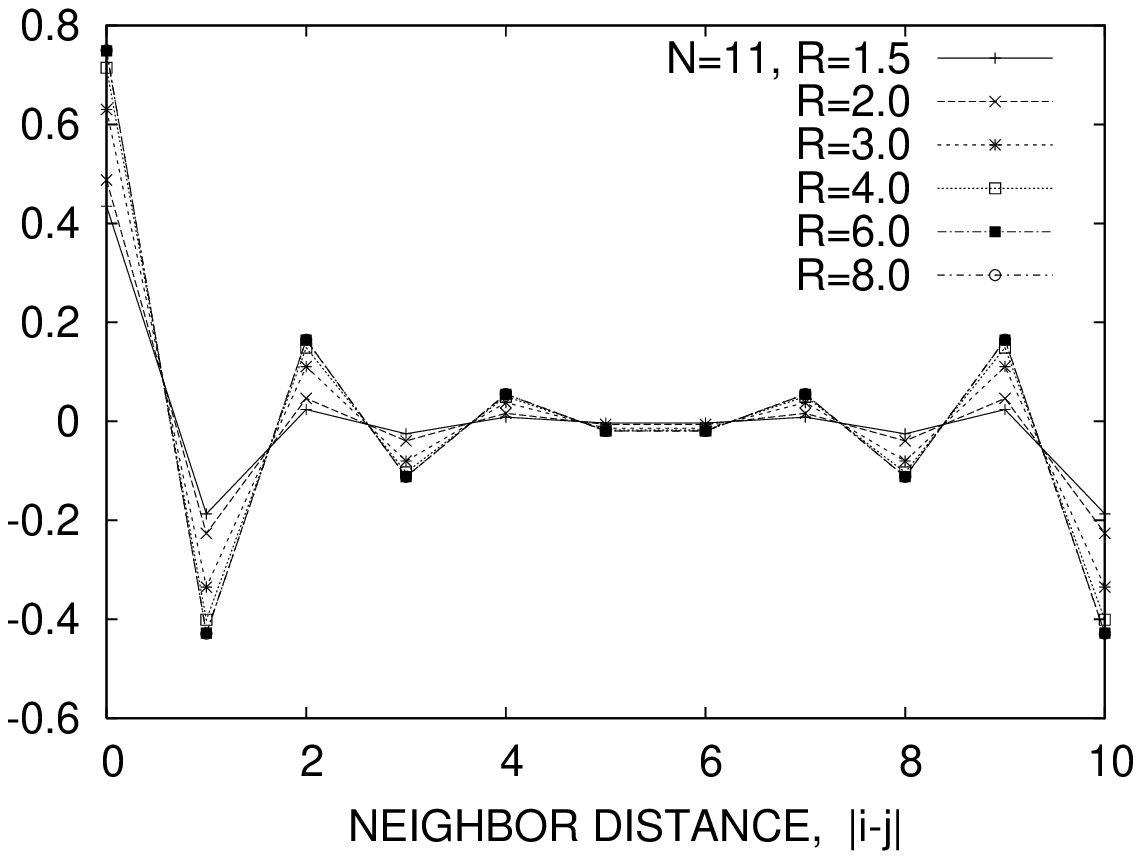}}
\put(1,16){\rotatebox{90}{\footnotesize 
  $\averaa{\mathbf{S}_i\cdot\mathbf{S}_j}$}}
\put(51,16){\rotatebox{90}{\footnotesize 
  $\averaa{\mathbf{S}_i\cdot\mathbf{S}_j}$}}
\put(8,30.5){\small $(a)$}
\put(58,30.5){\small $(b)$}
\end{picture}
\caption{Parity effect on spin ordering: spin--spin correlations
  for nanochains of $N=10$ $(a)$ and $N=11$ $(b)$ atoms. The values
  of the interatomic distance $R$ are specified in the atomic units 
  ($a_0=0.529\mbox{ \AA}$).}
\label{ssfig}
\end{figure}

We now discuss the spin--spin correlations 
$\averaa{\mathbf{S}_i\cdot\mathbf{S}_j}$ in the system ground state as 
a~function of the discrete neighbor distance $\abs{i\!-\!j}$ and $R$, as 
presented in Fig.\ \ref{ssfig}. The effect of spin frustration in the 
half--filled case ($N_e=N$) is remarkable for large $R$, where the 
quasi--long range order for $N=10$ (\emph{cf.}\ Fig.\ \ref{ssfig}a), 
indicating the power--law decay of $\averaa{\mathbf{S}_i\cdot\mathbf{S}_j}$
for the Heisenberg spin chain, disappears for $N=11$ (\emph{cf.}\ Fig.\ 
\ref{ssfig}b), where we observe a fast, exponential decay.
For small values of $R$, however, the effect is weaker, since the
antiferomagnetic order is reduced by charge fluctuations \cite{ryspa}.
We also observe, that the values of the spin gap (not shown) are significantly 
higher for odd $N$ in the large--$R$ range, what can be explained by the fact, 
that the ground--state energy of the Heisenberg antiferromagnet is of the
order $E_G\sim JS(S+1)$, where $J=4t^2/(U-K)$ is the kinetic--exchange coupling
parameter and $S$ is the total spin value. One can expect now, that the
following inequality is satisfied
$E_G^{S=\frac{3}{2}}-E_G^{S=\frac{1}{2}}>E_G^{S=1}-E_G^{S=0}$, where
the left-- and the right--hand sides represent the spin gap for the 
\emph{odd}-- and the
\emph{even}--$N$ systems, respectively (both at the minimal--spin 
configuration). The detailed behavior of the system spin, as well as the charge
and the optical gaps will be discussed elsewhere.
In the remaining part of this paper we focus on the parity effect for
the Fermi--Dirac distribution function and the transport properties.

\begin{figure}[!t]
\begin{minipage}[t]{0.48\textwidth}
\includegraphics[width=\textwidth]{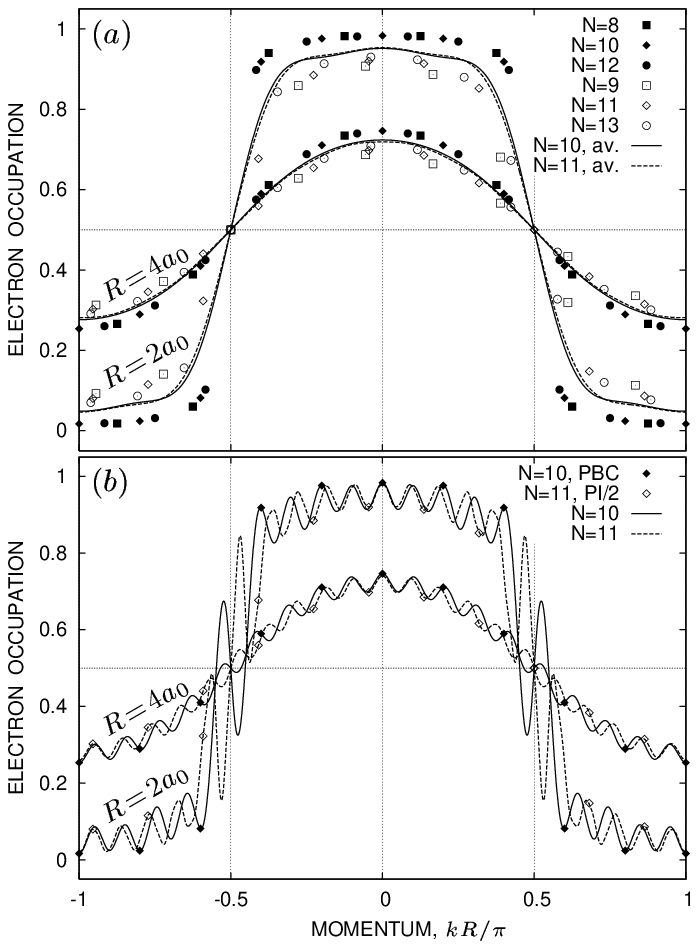}
\caption{Electron momentum distribution for chains of $N=8\div 13$ atoms:
$(a)$ datapoints for \emph{optimal} boundary conditions (BC\textit{s}) and 
the sample $n(k)$ curves averaged over BC\textit{s} (\emph{solid} lines); 
$(b)$ the original $n(k_q(\phi))$ functions used for the averaging.}
\label{nks}
\end{minipage}
\hfil
\begin{minipage}[t]{0.45\textwidth}
\includegraphics[width=\textwidth]{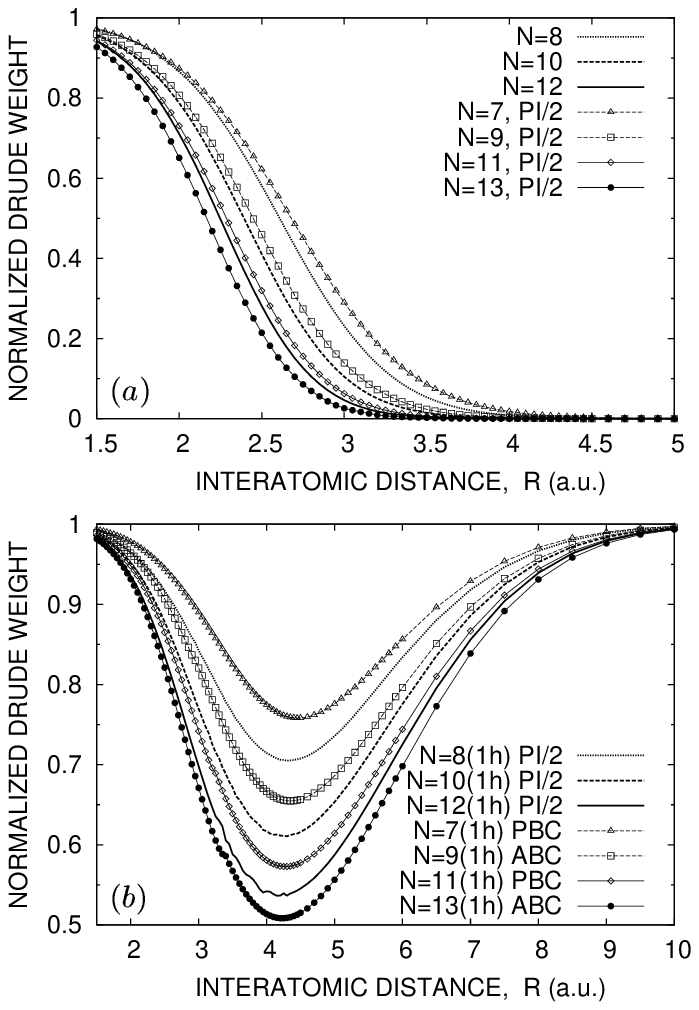}
\caption{Normalized Drude weight for nanochains in the \emph{half--filled} 
case $(a)$, and for a system with a single hole $(b)$. The \emph{optimal} 
boundary conditions are specified for each curve.}
\label{sigd}
\end{minipage}
\end{figure}

The electron momentum distribution for nanochains with \emph{optimal} 
BC\textit{s} is shown
in Fig.\ \ref{nks}a. The discrete momenta, corresponding to the solutions
of the single--particle part of the Hamiltonian (\ref{hameff}) for a finite 
$N$, are given by
\begin{equation}
\label{kqph}
  k_q(\phi)=\frac{2\pi q-\phi}{N},\ \ \ \  0\leqslant q<N.
\end{equation}
The \emph{optimal} BC\textit{s}, corresponding to the minimal 
ground--state energy $E_G$, are realized for $\phi=0$ when $N=4n+2$ 
(\emph{periodic} BC), $\phi=\pi$ when $N=4n$ (\emph{antiperiodic} BC)
and $\phi=\pi/2,3\pi/2$ when $N$ is odd \cite{nakano}.
A basic analysis of Eq.\ (\ref{kqph}) shows, that for the optimal BC\textit{s},
the Fermi momentum state $k_F=\pi/2R$ is newer reached for even $N$, whereas 
for odd $N$ it happens, for a single value of $q$. 
This circumstance has tremendous
implications for electronic structure of a nanochain, however, of almost
does not effect its transport properties, as we show in the end part of this
paper.

The remarkable feature of these results,
is that the datapoints for different but even $N$ (\emph{cf.}\ full symbols
in Fig.\ \ref{nks}a) locate smoothly on the 
universal curve (for each $R$), when optimal BC\textit{s} are applied. 
This is not the case for odd $N$ (\emph{cf.}\ open symbols in Fig.\ 
\ref{nks}a), when the systematic dependence on $N$ suggests quite slow
convergence to the even--$N$ results with the increasing $N$. 
However, the discussion of $N\rightarrow\infty$ limit is beyond the scope
of this paper, since we concentrate here on nanochains.
To analyze such systems in detail, we have displayed in Fig.\ \ref{nks}b 
the continuous electron momentum distribution obtained for the dense set
of $k_q(\phi)$ defined by Eq.\ (\ref{kqph}), when $\phi\in\langle 0,2\pi)$.
The datapoints corresponding to optimal BC for $N=10$ and $11$ are also 
presented to show, they are situated close to the different local extrema of 
$n(k_q(\phi))$ (e.g.\ maxima for even $N$ and minima for odd, or 
\emph{vice versa}).
Except of different frequency of internal oscillations (equal to $2NR$),
the $n(k_q(\phi))$ functions for $N=10$ and $11$ looks almost identically.
This is the reason, why various physical properties of small clusters are
often averaged over BC, particularly in 2D \cite{poilb}. 
We also perform such averaging to obtain almost size--independent $n(k)$ 
functions, drawn again in Fig.\ \ref{nks}a (the details of the averaging 
procedure will be published elsewhere). 
However, the elimination of the internal oscillations for a given $N$ 
may only be considered as an approximation of $N\rightarrow\infty$ scaling
procedure, and in the case of momentum distribution $n(k)$ seems less
accurate then fitting the Luttinger--liquid formulas to even $N$ data,
which we proposed before \cite{ryspa}. 
Nevertheless, the common nature of either original $n(k_q(\phi))$ or averaged
$n(k)$ functions for both even and odd $N$, illustrated in Fig.\ \ref{nks},
helps to understand why the chain parity does not effect its Drude weight
even for optimal boundary conditions, when the structure of the momentum space
is significantly different. 

The \emph{normalized} Drude weight, shown in Fig.\ \ref{sigd}, is defined in 
the standard manner \cite{shamil}
\begin{equation}
\label{drudew}
  D=-\frac{1}{\aver{T}}
  \left.\frac{\partial^2E_G}{\partial\phi^2}\right|_{\phi=\phi_{\min}},
\end{equation}
where $\averaa{T}$ is the average kinetic energy and $\phi_{\min}$ denotes 
optimal BC\textit{s}.
In the half--filled case $N_e=N$ (\emph{cf.}\ Fig.\ \ref{sigd}a) Drude
weight gradually decrease with $N$, as we have shown for even $N$ \cite{ryspa}.
The most interesting feature of these results is, that the curves for odd $N$
fits smoothly between those for even $N$, with very weak parity effect
(totally incomparable with that present in the charge gap and electron
momentum distribution, when optimal BC are applied).
This observation can be understood when we take into account, that the Drude
weight defined by Eq.\ \ref{drudew} is the \emph{integral} quantity,
involving the summation over all the excited states of the Hamiltonian 
Eq.\ (\ref{hameff}), so it cannot be determined only by the electronic 
structure near the Fermi points, particularly for a small system.

The parity effect on Drude weight disappears for the system with
a single hole ($N_e=N-1$, \emph{cf.}\ Fig.\ \ref{sigd}b), in which
the magnetic frustration is absent.
For this case, the Drude weight evolution with $R$ is very interesting.
In the weak--correlation range ($R/a_0\lesssim 2$) the chain shows a 
highly--conducting behavior for each $N$. Next, in the intermediate range
($R/a_0=4\div 5$) the Drude weight decrease rapidly with $N$, indicating an
insulating (Mott--Hubbard) state in the large--$N$ limit. 
In the strongly--correlated range ($R/a_0\sim 10$) the Drude weight
approaches again its maximal value $D=1$.
Such a behavior can be explained, when we analyze the situation in two steps:
\emph{First,} for low values of $R$, the bandwidth--to--interaction ratio
is small, and the system with a single hole does not differ significantly
from a half--filled one. This is why in both cases Drude weight decreases
gradually with both $N$ and $R$, as the tunneling amplitude through the barrier
of a finite width. 
\emph{Second,} for the largest values of $R$, the system can be described
by an effective $t-J$ model \cite{spatj} with a coupling constant
$J=4t^2/(U-K)\ll |t|$  (where $K\equiv K_{j,j+1}$ denotes the nearest
neighbor Coulomb repulsion), which corresponds to an asymptotically--free 
hole motion.
Then, it become clear that in the intermediate range the Drude weight has 
to suppressed, what can be interpreted in terms of a partially localized 
spin--ordered state. 
It would be very interesting to test experimentally this result,
possibly for a mesoscopic atomic ring.

In summary, we have shown that a nanochain parity effect strongly
its electronic structure and momentum distribution, and that the effects
are opposite for these two principal characteristics. Namely, the presence
of a discrete momenta in the Fermi point reduce significantly the finite--size
effects on the system charge--gap, but amplifies them in the case of momentum
distribution. On the other hand, the parity effect is weak, or even absent
in the case if system transport properties. 
Additionally, an interesting crossover behavior has been identified for
the chain with a single hole, for which the quantum--liquid regions are 
separated by a partly localized state.

The support from the Polish Science Foundation (FNP) is acknowledged.



\begin{thebibliography}{00}
\bibitem{spary} 
J.\ Spa{\l}ek, \emph{et al.}, 
\textit{Phys.\ Rev.\ B} \textbf{61}, 15676 (2000); 
A.\ Rycerz, J.\ Spa\l ek, \emph{ibid.} \textbf{63}, 073101 (2001); 
\textbf{65}, 035110 (2002). 

\bibitem{genbc}
R.\ Jullien, R.\ M.\ Martin, 
\textit{Phys.\ Rev.\ B} \textbf{26}, 6173 (1982);
B.\ Fourcade, G.\ Sproken, 
\emph{ibid.} \textbf{29}, 5096 (1984);
R.\ M.\ Fye, \emph{et al.}, 
\emph{ibid.} \textbf{44}, 6909 (1991);
K.\ Ro\'{s}ciszewski, A.\ M.\ Ole\'{s},
\textit{J.\ Phys.\ Cond.\ Mat.} \textbf{5}, 7289, (1993).

\bibitem{nakano}
E.\ H.\ Lieb, 
\textit{Phys.\ Rev.\ Lett.} \textbf{73}, 2158 (1994);
F.\ Nakano, 
\textit{J.\ Phys. A} \textbf{33}, 5429 (2000);
\emph{ibid.} \textbf{37}, 3979 (2004).

\bibitem{sparev} 
J.\ Spa{\l}ek, \emph{et al.}, 
\textit{Acta Phys.\ Polon.\ B} \textbf{31}, 2879 (2000); 
\emph{ibid.} \textbf{32}, 3189 (2001);
in \textit{Concepts in Electron Correlation}, 
Proc.\ of the NATO Adv.\ Res.\ Workshop, eds A.C.\ Hewson and V.\ Zlati\'{c},
pp.\ 257--268, Kluwer, Dordrecht (2003).

\bibitem{ryspa}
A.\ Rycerz, J.\ Spa{\l}ek, 
\textit{Eur.\ Phys.\ J.\ B} \textbf{40}, 153 (2004).

\bibitem{poilb}
D.\ Poilblanc, 
\textit{Phys.\ Rev.\ B} \textbf{44}, 9562 (1991).

\bibitem{davidson}
E.\ R.\ Davidson, 
\textit{J.\ Comput.\ Phys.} \textbf{17}, 87 (1975).

\bibitem{shamil}
B.S.\ Shastry and B.\ Sutherland 
\textit{Phys.\ Rev.\ Lett.} \textbf{65}, 243 (1990);
J.A.\ Millis and S.N.\ Coppersmith,
\textit{Phys.\ Rev.\ B} \textbf{42}, 10807 (1990);
D.\ G\'{o}ra, K.\ Ro\'{s}ciszewski, A.M.\ Ole\'{s},
J.\ Phys.\ Condensed Matter \textbf{10}, 4755 (1998).

\bibitem{spatj}
P.\ W.\ Anderson, \textit{Phys. Rev.} \textbf{115}, 2 (1959); 
W.\ F.\ Brinkman and T.\ M.\ Rice, 
\textit{Phys.\ Rev.\ B} \textbf{2}, 1324 (1970); 
K.\ A.\ Chao, J.\ Spa{\l}ek, and A.\ M.\ Ole\'s, \textit{J.\ Phys.\ C}
\textbf{10}, L271 (1977).
\end{thebibliography}
\end{document}